\documentclass[11pt,oneside]{article}
\usepackage[T1]{fontenc}
\usepackage[OT4]{fontenc}
\usepackage{a4}
\usepackage{amsfonts}
\usepackage{amsmath}
\usepackage{amssymb}
\usepackage{amsxtra}
\usepackage{amsthm}

\let\subsetnew\subset
\let\nsubequal\nsubseteq
\let\subequal\subseteq
\usepackage{mathabx}

\usepackage[dvips]{graphicx}
\usepackage{pstricks}
\usepackage{color}
\usepackage{bezier}
\usepackage{epic}
\usepackage{eepic}

\newtheorem{thm}{Theorem}
\newtheorem{cor}{Corollary}

\newtheorem{lem}{Lemma}
\newtheorem{Problem}{Problem}
\theoremstyle{remark}

\newtheoremstyle{cited}%
  {3pt}
  {3pt}
  {\itshape}
  {}
  {\bfseries}
  {.}
  {.5em}
  {\thmname{#1} \thmnumber{#2} \thmnote{\normalfont#3}}

\theoremstyle{cited}

\title{\bf Hangable graphs}

\author{Mateusz Miotk, Jerzy Topp}

\begin{document}

\maketitle
\begin{center}{\small Faculty of Mathematics, Physics and Informatics }\\ {\small  University of Gda\'nsk, 80-952 Gda\'nsk, Poland} \end{center}

\noindent {\bf Abstract}. Let $G=(V_G,E_G)$ be a connected graph. The distance
$d_G(u,v)$ between vertices $u$ and $v$ in $G$ is the length of a shortest $u-v$ path
in $G$. The eccentricity of a vertex $v$ in $G$ is the integer $e_G(v)= \max\{ d_G(v,u)
\colon u\in V_G\}$. The diameter of $G$ is the integer $d(G)= \max\{e_G(v)\colon
v\in V_G\}$. The periphery of a~vertex $v$ of $G$ is the set $P_G(v)= \{u\in V_G\colon
d_G(v,u)= e_G(v)\}$, while the periphery of $G$ is the set $P(G)= \{v\in V_G\colon
e_G(v)=d(G)\}$. We say that graph $G$ is hangable if $P_G(v)\subequal
P(G)$ for every vertex $v$ of $G$. In this paper we prove that every block
graph is hangable and discuss the hangability of products of graphs.\\


\section{Introduction}

\begin{figure*}[!h]\begin{center} \special{em:linewidth 0.4pt}
\unitlength 0.4ex \linethickness{0.4pt} \begin{picture}(60,45)\put(-25,5){
\path(0,15)(15,0)(30,15)(15,30)(0,15)\path(30,15)(50,15)\path(15,0)(15,30)
\put(0,15){\circle*{1.5}}\put(15,0){\circle*{1.5}}\put(15,30){\circle*{1.5}}
\put(30,15){\circle*{1.5}}\put(50,15){\circle*{1.5}}
\put(-1.2,9){${}^{a}$}\put(14,30){${}^{b}$}\put(13.5,-7){${}^{d}$}
\put(29,9){${}^{c}$}\put(49,9){${}^{e}$}\put(-10,29){${}^{G}$}}
\put(60,5){\path(0,15)(15,0)(30,15)(15,30)(0,15)\path(15,0)(15,30)
\put(0,15){\circle*{1.5}}\put(15,0){\circle*{1.5}}\put(15,30){\circle*{1.5}}
\put(30,15){\circle*{1.5}}\put(-1.2,9){${}^{a}$}\put(14,30){${}^{b}$}
\put(13.5,-7){${}^{d}$}\put(29,9){${}^{c}$}\put(-10,29){${}^{H}$}}
\end{picture}\caption{A hangable graph $G$ and a non-hangable graph $H$}
\label{hangable-no-hangable} \end{center} \end{figure*}
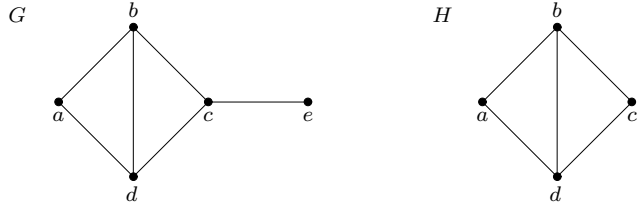

\noindent We use \cite{BuckleyHarary} and \cite{GrossYellen} for basic terminology
and notations.  Let $G=(V_G,E_G)$ be a connected graph. The distance $d_G(u,v)$
between vertices $u$ and $v$ in $G$ is the length of a shortest $u-v$ path in $G$.
The eccentricity $e_G(v)$ of a vertex $v$ in $G$ is the distance from $v$ to
a vertex farthest from $v$, that is $e_G(v)= \max\{ d_G(v,u)\colon u\in V_G\}$.
The diameter $d(G)$ of $G$ is the maximum eccentricity of the vertices of $G$.
It follows from these definitions that $d(G)= \max\{d_G(u,v)\colon u, v \in V_G\}$.
The periphery of  vertex $v$ of $G$ is the set $P_G(v)$ of the vertices farthest
from $v$, $P_G(v)= \{u\in V_G\colon d_G(v,u)= e_G(v)\}$, whereas the periphery
$P(G)$ of  graph $G$ is the set of vertices having the maximum eccentricity in $G$,
that is $P(G)= \{v\in V_G\colon e_G(v)=d(G)\}$. A connected graph $G$ is said to
be self-centered if  $P(G)=V_G$. We say that graph $G$ is hangable if $P_G(v)
\subequal P(G)$ for every vertex $v$ of $G$. Note that graph $G$ is hangable if
and only if it has the property: for every three vertices $v$, $u$, and $w$,  if $u$
is a vertex farthest from $v$, and $w$ is a vertex farthest from $u$, then the distance $d_G(u,w)$ is the diameter of $G$. Every self-centered graph $G$ is hangable since
$P(G)=V_G$ and therefore $P_G(v)\subequal P(G)$ for every vertex $v$ of $G$.
In particular, every complete graph $K_n$ is hangable.  Similarly, every cycle $C_n$
and every $n$-cube $Q_n$ are hangable. Every complete bipartite graph $K_{m,n}$ is
also a hangable graph, as is easy to check. Graph $G$ in Fig. \ref{hangable-no-hangable}
is hangable since $P_G(v)\subequal P(G)$ for every vertex $v$ of $G$ ($P_G(a)
= P_G(b)= P_G(d)=\{e\}\subequal \{a, e\}=P(G)$ and $P_G(c)= P_G(e)= \{a\}\subequal
\{a, e\}=P(G)$). Graph $H$ in Fig. \ref{hangable-no-hangable} is not hangable
as $P_H(v)\nsubequal P(H)$ for some vertex $v$ of $H$ ($P_H(b)= P_H(d)=\{a, c, d\}\not\subequal\{a, d\} =P(H)$).

In this paper we prove that every block graph is hangable and
discuss the hangability of three products of  graphs.

\section{Hangability of block graphs}

We begin by recalling that a block graph is a connected graph in which every block
(i.e., every maximal 2-connected subgraph) is a~complete graph.
Hangability of block graphs is ascertained by the following theorem.

\begin{thm} \label{hangable-block} Every block graph $G$ is hangable. \end{thm}

\noindent {\bf Proof.} 
Let $v$ be any vertex of $G$, and let $G_v$ be the graph $G$ rooted (hanged) at $v$.
The proof will be complete if we show that $P_G(v)\subequal P(G)$.
To prove this it suffices to show that if $u\in P_G(v)$ and $x$ and $y$ are vertices
of $G$, then $d_G(u,x)\ge d_G(x,y)$ or $d_G(u,y)\ge d_G(x,y)$.

This is obvious if $x=y$ or $x$ and $y$ belong to the shortest $v-u$ path in $G_v$.
Thus assume that at most one of the vertices $x$ and $y$  belong to the shortest
$v-u$ path in $G_v$, and let $a_x$ ($a_y$, resp.) be the youngest common ancestor
of the vertices $u$ and $x$ ($u$ and $y$, resp.) in the rooted graph $G_v$.

\begin{figure*}[!h]\begin{center} \special{em:linewidth 0.4pt}
\unitlength 0.4ex \linethickness{0.4pt} \begin{picture}(110,85)
\put(1,0){\put(-35,5){\path(0,0)(0,70)\put(0,0){\circle*{1.5}}\put(0,50){\circle*{1.5}}\put(0,60){\circle*{1.5}}
\put(0,70){\circle*{1.5}}\put(10,5){\circle*{1.5}}\put(-9,72){${}^{(\rm a)}$}
\path(0,50)(10,5)\put(-4,-3){${}^{u}$}\put(2,67){${}^{v}$}
\put(2,57){${}^{a_x=x}$}\put(8.75,-1){${}^{y}$}\put(2,47){${}^{a_y}$}}

\put(3,5){
{\path(0,0)(0,70)\put(0,0){\circle*{1.5}}\put(0,50){\circle*{1.5}}\put(0,30){\circle*{1.5}}
\put(0,41){\circle*{1.5}}\put(5,42){\circle*{1.5}}\multiput(0,41)(1,0.2){5}{\circle*{0.3}}
\put(0,70){\circle*{1.5}}\put(10,5){\circle*{1.5}}\put(-9,72){${}^{(\rm b)}$}
\path(0,50)(5,42)(10,5)\put(-4,-3){${}^{u}$}\put(2,67){${}^{v}$}
\put(-12.9,27){${}^{a_x=x}$}\put(8.75,-1){${}^{y}$}\put(2,47){${}^{a_y}$}}}
\put(42,5){\path(0,0)(0,70)\put(0,0){\circle*{1.5}}\put(0,50){\circle*{1.5}}
\put(-9,72){${}^{(\rm c)}$}\put(0,70){\circle*{1.5}}\put(0,30){\circle*{1.5}}
\put(10,5){\circle*{1.5}}\put(10,25){\circle*{1.5}}
\path(0,50)(10,25)\path(10,5)(0,30)\put(-4,-3){${}^{u}$}\put(2,67){${}^{v}$}
\put(8.5,-1){${}^{x}$}\put(8.75,19){${}^{y}$}\put(2,47.5){${}^{a_y}$}\put(-6,27.5){${}^{a_x}$}}
\put(78,5){\path(0,0)(0,70)\put(0,0){\circle*{1.5}}\put(0,50){\circle*{1.5}}
\put(-9,72){${}^{(\rm d)}$}\put(0,70){\circle*{1.5}}\put(10,30){\circle*{1.5}}
\put(10,5){\circle*{1.5}}\put(20,5){\circle*{1.5}}\path(0,50)(5,42)(10,30)(15,22)(20,5)
\path(10,5)(10,30)\put(-4,-3){${}^{u}$}\put(2,67){${}^{v}$}
\put(0,41){\circle*{1.5}}\put(5,42){\circle*{1.5}}\put(10,21){\circle*{1.5}}\put(15,22){\circle*{1.5}}
\put(8.5,-1){${}^{x}$}\put(18.5,-1){${}^{y}$}\put(1,48.5){${}^{a_x=a_y}$}\put(11,28.5){${}^{a_{xy}}$}
\multiput(0,41)(1,0.2){5}{\circle*{0.3}}\multiput(10,21)(1,0.2){5}{\circle*{0.3}}
\put(-4,38){${}^{a}$}\put(6.5,39){${}^{b}$}}
\put(118,5){\path(10,0)(10,70)\put(10,0){\circle*{1.5}}\put(10,50){\circle*{1.5}}
\put(3,72){${}^{(\rm e)}$}\put(10,70){\circle*{1.5}}\put(0,5){\circle*{1.5}}
\put(20,5){\circle*{1.5}}\path(0,5)(5,41)(10,50)(15,41)(20,5)\put(6,-3){${}^{u}$}
\put(12,67){${}^{v}$}\put(10,40){\circle*{1.5}}\put(5,42){\circle*{1.5}}
\put(15,42){\circle*{1.5}}\multiput(10,40)(1,0.4){5}{\circle*{0.3}}
\multiput(10,40)(-1,0.4){5}{\circle*{0.3}}\multiput(5,42)(1,0){10}{\circle*{0.3}}
\put(1.5,39.5){${}^{c}$}\put(11.1,35.5){${}^{a}$}\put(16.5,39.5){${}^{d}$}
\put(-1.5,-1){${}^{x}$}\put(18.75,-1){${}^{y}$}\put(11,48.5){${}^{a_{xy}=a_x=a_y}$}}}
\end{picture}\caption{Block graphs rooted (hanged) at the vertex $v$}\label{picture1111} \end{center} \end{figure*}
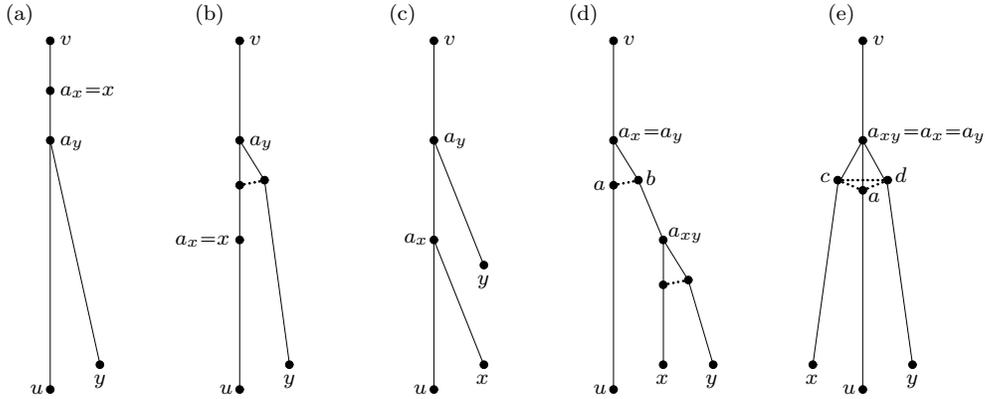

First assume that one of the vertices $x$ and $y$, say $x$, belongs to the shortest
$v-u$ path in $G_v$. If $x$ $(=a_x)$ is not younger than $a_y$ in $G_v$ (see Fig. \ref{picture1111}\,(a)), then the choice of $u$ implies that $d_G(x,y) \le d_G(x,u)$.
If $x$ is younger than $a_y$ (Fig. \ref{picture1111}\,(b)), then $x$ belongs to the shortest $u-y$ path and therefore $d_G(u,y)= d_G(u,x)+d_G(x,y)\ge d_G(x,y)$.

Now assume that none of the vertices $x$ and $y$ belongs to the shortest $v-u$ path in $G_v$. If $a_x$ is younger than $a_y$ in $G_v$ (Fig. \ref{picture1111}\,(c)), then $d_G(u,y)=d_G(u,a_x)+d_G(a_x,y)\ge d_G(x,a_x)+ d_G(a_x,y)= d_G(x,y)$ as $d_G(a_x,u)\ge d_G(a_x,x)$. Similarly, $d_G(u,x)\ge d_G(x,y)$ if $a_y$ is younger than $a_x$ in $G_v$. Thus assume $a_x=a_y$. In this case let $a_{xy}$ be the youngest common
ancestor of the vertices $x$ and $y$ in $G_v$. We consider two possible subcases: $a_{xy} \neq a_x=a_y$, and $a_{xy}=a_x=a_y$.

If $a_{xy}\neq a_x=a_y$ and $a_{xy}\in \{x,y\}$, say $a_{xy}=x$, then $x$ belongs
to the shortest $u-y$ path and therefore $d_G(u,y)= d_G(u,x)+d_G(x,y)\ge d_G(x,y)$.
Thus assume $a_{xy}\neq a_x=a_y$ and $a_{xy}\not\in \{x,y\}$ (Fig. \ref{picture1111}\,(d)). Let $a$ and $b$ be the neighbours of $a_x$
which belong to the shortest $u-a_x$ and $a_{xy}-a_x$ paths, respectively.
Then, since $d_G(a,u)\ge d_G(b,y)$, we have $d_G(u,x)\ge d_G(u,a)+d_G(b,x)+1\ge d_G(y,b)+d_G(b,x)+1\ge d_G(y,a_{xy})+ d_G(a_{xy},x)+1 \ge d_G(y,x)+1\ge d_G(y,x)$.

Finally assume that $a_{xy}=a_x=a_y$. Let $a$, $c$, and $d$ be the neighbours of
$a_x$ which belong to the shortest $u-a_x$, $x-a_x$, and $y-a_x$ paths, respectively, Fig. \ref{picture1111}\,(e). Since $G$ is a block graph, the subgraph of $G$ induced
by the vertices $a$, $c$, and $d$ has either at most one or three edges. In the first case we may assume that $ac$ is not an edge of $G$. Then $d_G(u,x)= d_G(u,a_x)+d_G(a_x,x) \ge d_G(y,a_x)+d_G(a_x,x) \ge d_G(y,x)$. In the second case $ac$, $ad$, and $cd$ are edges of $G$, and then the choice of $u$ implies that $d_G(a,u)\ge d_G(c,x)$ and therefore $d_G(u,x)\ge d_G(u,a)+1+d_G(c,x)\ge d_G(y,d)+1+d_G(c,x)=d_G(y,x)$.

This completes the proof of the theorem.

Since every tree is a block graph, from Theorem \ref{hangable-block} we immediately have the following corollary.

\begin{cor} Every tree is a hangable graph. \end{cor}

\section{Hangability of product graphs}

\begin{figure*}[!h]\begin{center} \special{em:linewidth 0.4pt}
\unitlength 0.6ex \linethickness{0.4pt} \begin{picture}(60,35)\put(2,5){
\path(55,0)(0,0)(15,10)(30,0)\path(0,0)(4,15)(-4,15)(0,0)\put(0,0){\put(0,0){\circle*{1.5}}%
\put(4,15){\circle*{1.5}}\put(-4,15){\circle*{1.5}}}
\put(30,0){\path(0,0)(4,15)(-4,15)(0,0)\put(0,0){\circle*{1.5}}%
\put(4,15){\circle*{1.5}}\put(-4,15){\circle*{1.5}}}
\put(55,0){\path(0,0)(4,15)(-4,15)(0,0)\put(0,0){\circle*{1.5}}%
\put(4,15){\circle*{1.5}}\put(-4,15){\circle*{1.5}}}
\put(15,10){\path(0,0)(4,15)(-4,15)(0,0)\put(0,0){\circle*{1.5}}%
\put(4,15){\circle*{1.5}}\put(-4,15){\circle*{1.5}}}}
\end{picture}\caption{The corona $(K_1+(K_1\cup K_2))\circ K_2$}
\label{przyklad-korony} \end{center} \end{figure*}
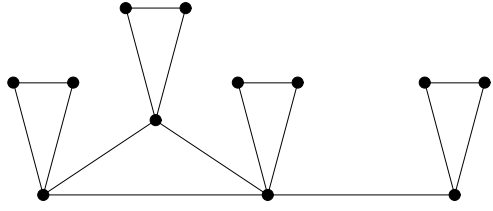

We now turn our attention to the hangability of coronas of graphs.
Let $G$ and $H$ be two graphs. The corona of G and H, denoted by $G \circ H$, is the graph with vertex set $V_G \cup (V_G\times V_H)$ and edge set $E_G \cup \bigcup_{v \in V_G}\{v(v,x)\colon x \in V_H\} \cup \bigcup_{v \in V_G}\{(v,x)(v,y)\colon xy \in E_H\}$. In Fig. \ref{przyklad-korony}, the corona $G\circ H$ is shown in which $G=
K_1+(K_1\cup K_2)$ and $H=K_2$.

Basic properties of distances, diameters and sets of peripheral vertices of coronas
are summarized in the following lemma.

\begin{lem}\label{LemcoronaKn} If $G\circ H$ is the corona of graphs $G$ and $H$, where $G$ is a  connected  graph of order at least two, then: \begin{enumerate} \item[$(1)$] $d_{G\circ H}((u,x),(v,y))= d_G(u,v)+2$ if $u, v\in V_G$ and $x, y\in V_H$,\\[1ex] $d_{G\circ H}((u,x),v)= d_G(u,v)+1$ if $u, v\in V_G$ and $x\in V_H$,\\[1ex] $d_{G\circ H}(u,v)= d_G(u,v)$ if $u, v\in V_G$; \item[$(2)$] $d(G \circ H) = d(G) + 2$; \item[$(3)$] $P_{G\circ H}(u)= P_{G\circ H}((u,x))= P_G(u)\times V_H$ if $u\in V_G$ and $x\in V_H$;
\item[$(4)$] $P(G \circ H) = P(G) \times V_H$. \end{enumerate}\end{lem}

\noindent {\bf Proof.} Since (1), (2), and (3) are obvious, we only prove (4).
For this  the inclusions  $P(G \circ H) \subequal P(G) \times V_H$ and  $P(G) \times V_H \subequal P(G \circ H)$ have to be proved.

Assume first that $a \in P(G \circ H)$. Let $b$ be a vertex of $G\circ H$ for which
$d_{G\circ H}(a,b)= d(G\circ H)$. From the definition of $G\circ H$ (and by (1)) it is obvious that $a=(v,x)$ and $b=(u,y)$ for some vertices $v$ and $u$ of $G$ and some
vertices $x$ and $y$ of $H$. Now, by (1) and (2) it follows
that $d(G)+2 = d(G\circ H)= d_{G\circ H}(a,b)=d_{G\circ H}((v,x),(u,y))= d_G(v,u)+2$. Thus $d_G(v,u)=d(G)$ and this implies that $v\in P(G)$. Therefore $a=(v,x)\in P(G)\times
V_H$ and, consequently, $P(G \circ H) \subequal P(G) \times V_H$.

Assume now that $(v,x)\in P(G)\times V_H$. Let $u\in V_G$ be such that $d_G(v,u)= d(G)$. Then, by (1), $d_{G\circ H}((v,x),(u,x))= d_G(v,u)+2= d(G)+2$. From this and from (2) it follows that $d_{G\circ H}((v,x),(u,x))= d(G\circ H)$ and therefore $(v,x)\in
P(G\circ H)$. This proves that $P(G) \times V_H \subequal P(G \circ H)$.\\

Now we are ready to prove a necessary and sufficient condition for a corona of graphs
to be hangable.

\begin{thm} If $G$ is a connected graph of order at least two and $H$ is any graph, then
the corona $G \circ H$ is a hangable graph if and only if  $G$ is hangable.\end{thm}

\noindent {\bf Proof.} Let $G$ be a hangable graph. Let $u$ and $x$ be any vertex of $G$ and $H$, respectively. Then $P_G(u)\subequal P(G)$ and consequently
$P_G(u)\times V_H \subequal P(G)\times V_H$. Thus, by statements (3) and (4) of Lemma \ref{LemcoronaKn}, $P_{G\circ H}(u) \subequal P(G\circ H)$ and $P_{G\circ H}((u,x))
\subequal P(G\circ H)$. This proves that graph $G\circ H$ is hangable.

Assume now that  graph $G\circ H$ is hangable. Then $P_{G\circ H}(u)\subequal
P(G\circ H)$ and $P_{G\circ H}((u,x))\subequal P(G\circ H)$, where $u\in V_G$ and $(u,x)\in V_G\times V_H$ are two possible types of vertices of $G\circ H$. Now, since
$P_{G\circ H}(u)=P_{G\circ H}((u,x))=P_G(u)\times V_H$ and $P(G\circ H)= P(G)\times
V_H$ (by statements (3) and (4) of Lemma \ref{LemcoronaKn}), it follows that
 $P_G(u)\times V_H \subequal P(G)\times V_H$. Consequently, $P_G(u)\subequal P(G)$
 and this proves that $G$ is hangable. \\

Let $G$ and $H$ be graphs. The cartesian product of  $G$ and $H$, denoted by $G \square H$, is the graph with vertex set $V_G \times V_H$ and where two vertices $(a,b)$ and $(c,d)$ are adjacent if and only if $ac \in E_G$ and $b=d$ or $a = c$ and $bd \in E_H$.
Distances,  eccentricities, diameters and sets of peripheral vertices of cartesian products of graphs are discussed in the next lemma.

\begin{lem} \label{Cartesian,dist,ecc,diam,per} If $G\square H$ is the cartesian product
of connected graphs $G$ and $H$, then: \begin{enumerate} \item[$(1)$] $d_{G \square H} ((a,b), (c,d)) = d_G(a,c) + d_H(b,d)$, if $(a,b),\, (c,d)\in V_{G\square H}$;
\item[$(2)$] $e_{G \square H}(a,b) = e_G(a) + e_H(b)$, if $(a,b)\in V_{G\square H}$;
\item[$(3)$] $d(G \square H) = d(G) + d(H)$; \item[$(4)$] $P_{G \square H}((a,b))$ = $P_G(a) \times P_H(b)$, if $(a,b)\in V_{G\square H}$; \item[$(5)$] $P(G \square H) = P(G) \times P(H)$.
\end{enumerate} \end{lem}

\noindent {\bf Proof.} The statements (1)-(3) without any proof were presented
 in \cite{BuckleyLewinter}. A formal proof of (1) was  given in
 \cite{HandbookProducts}. We prove the statements (2)--(5).

(2) Let us take vertices $a'\in P_G(a)$, $b'\in P_H(b)$, and $(c,d)\in P_{G\square H}((a,b))$. Then, by (1), we have \begin{center} $\begin{array}{rcl} e_G(a) + e_H(b) &=& d_G(a,a')+ d_H(b,b') = d_{G\square H}((a,b),(a',b'))\\[0.5ex] &\le & e_{G\square H}((a,b))= d_{G\square H}((a,b), (c,d))\\[0.5ex] &=& d_G(a,c)+d_H(b,d) \le e_G(a) + e_H(b)\end{array}$\end{center}  and this proves that $e_{G \square H}(a,b) = e_G(a) + e_H(b)$.

(3) Let $a,\, a',\, b,\, b',\, (c,d),\, (e,f)$ be vertices of $G$, $H$, and $G \square H$, where $d_G(a,a') = d(G)$, $d_H(b,b') = d(H)$, and $d_{G
\square H}((c,d),(e,f)) = d(G \square H)$, respectively. From the choice of vertices and by (1)  it follows that  \begin{center} $\begin{array}{rcl} d(G) + d(H)
&=& d_G(a,a') + d_H(b,b') = d_{G \square H}((a,b),(a',b')) \\[0.5ex] &\leq
& d(G \square H) = d_{G \square H}((c,d),(e,f)) = d_G(c,e) + d_H(d,f)\\[0.5ex]
&\leq & d(G) + d(H).\end{array}$ \end{center}
This implies that $d(G \square H) = d(G) + d(H)$.

(4) The definition of periphery of a vertex and properties (1) and (2) validate the following chain equalities \[\begin{array}{rcl} P_{G \square H}(a,b) &=& \{(x,y)\in V_{G\square H}\colon d_{G \square H}((a,b),(x,y))= e_{G \square H}(a,b)\}\\
&=& \{(x,y)\in V_{G\square H}\colon d_G(a,x)+d_H(b,y)= e_G(a)+e_H(b)\}\\
&=& \{(x,y)\in V_{G\square H}\colon d_G(a,x)=e_G(a)\,\,\mbox{and}\,\,
d_H(b,y)= e_H(b)\}\\ &=& \{(x,y)\in V_{G\square H}\colon x\in P_G(a)\,\,\mbox{and}
\,\, y\in P_H(b)\}\\ &=& P_G(a)\times P_H(b). \end{array}\]

(5) From the definition of periphery of a graph and from (2) and (3) it follows that \[\begin{array}{rcl} P(G\square H) &=& \{(a,b)\in V(G\square H)\colon e_{G\square H}((a,b))= d(G\square H)\}\\  &=& \{(a,b)\in V(G\square H)\colon e_G(a)+e_H(b)= d(G)+d(H)\}\\ &=& \{(a,b)\in V(G\square H)\colon e_G(a)= d(G)\,\, \mbox{and} \,\,e_H(b)= d(H)\}\\  &=& \{(a,b)\in V(G\square H)\colon a\in P(G) \,\, \mbox{and} \,\, b\in P(H)\}\\ &=& P(G)\times P(H). \end{array}\]

The following theorem specifies when the cartesian product of graphs is a~hangable graph and shows how to construct one hangable graph from other graphs.

\begin{thm} If $G$ and $H$ are connected graphs, then the cartesian product $G \square H$ is a hangable graph, if and only if $G$ and $H$ are hangable graphs.
\end{thm}

\noindent {\bf Proof.} Let $(a,b)$ be a vertex of $G\square H$.
Since $P_{G \square H}((a,b)) = P_G(a) \times P_H(b)$ (by Lemma \ref{Cartesian,dist,ecc,diam,per}\,(4)) and $P(G\square H)= P(G)\times P(H)$ (by Lemma \ref{Cartesian,dist,ecc,diam,per}\,(5)), the result follows from the equivalences \[\begin{array}{rcl} P_{G\square H}((a,b))\subequal P(G\square H) &\Leftrightarrow & P_G(a)\times P_H(b)\subequal P(G)\times P(H)\\[0.5ex] &\Leftrightarrow & P_G(a)\subequal P(G) \,\,\mbox{and}\,\, P_H(b)\subequal P(H).\end{array}\]

\begin{cor} The cartesian product $P_m\square  P_n$ is a hangable graph for every
paths $P_m$ and\, $P_n$. \end{cor}

The join of graphs $G$ and $H$, denoted $G+H$, is the graph obtained from the disjoint union of $G$ and $H$ by adding an edge between each vertex of $G$ and each vertex of $H$. The following theorem relates to the hangability of join graphs.

\begin{thm}\label{hangable-join} For graphs $G$ and $H$, the join $G+H$ is a hangable
graph if and only if either \begin{enumerate} \item[$(1)$] $G+H$ is a complete graph, or \item[$(2)$] at most one vertex of $G+H$ is adjacent to every other vertex of $G+H$.
\end{enumerate}\end{thm}

\noindent {\bf Proof.} Assume first that $G+H$ is a non-complete hangable graph.
Let $U$ be the set of all vertices, each of which is adjacent to every other vertex of $G+H$. It remains to show that $U$ has at most one element. Suppose that $U$ is non-empty. Then, since $e_{G+H}(x)=1$ for every $x\in U$ and $e_{G+H}(y)=2$ for every $y\in V_{G+H}-U$, it
follows that $P(G+H)=V_{G+H}-U$. Now, since $P_{G+H}(x)=V_{G+H}-\{x\}$ for every $x\in U$, the hangability of $G+H$ implies that $V_{G+H}-\{x\}$ is a subset of $V_{G+H}-U$
$(=P(G+H))$ and this clearly forces that $U$ has at most one element.

If $G+H$ is a complete graph, then $G+H$ is a hangable graph as we have already
observed. If no vertex of $G+H$ is adjacent to every other vertex of $G+H$, then
$e_{G+H}(x)=2$ for every $x\in V_{G+H}$. Therefore $P(G+H)=V_{G+H}$ and this
implies that $G+H$ is hangable. Finally assume that $G+H$ has exactly one vertex,
say $v_0$, which is adjacent to every other vertex of $G+H$. Then $e_{G+H}(v_0)=1$
and $e_{G+H}(x)=2$ for every $x\in V_{G+H}-\{v_0\}$. Consequently $P(G+H)
= V_{G+H}-\{v_0\}$ and now it is obvious that $G+H$ is a hangable graph since  $P_{G+H}(v_0)=V_{G+H}-\{v_0\} \subequal P(G+H)$ and
$P_{G+H}(x) \subsetnew P(G+H)$ for every $x\in V_{G+H}-\{v_0\}$.\\

The last theorem allows us to obtain  hangable graphs from other graphs.
Using the join operation, it is also easy to describe how to embed
(as an induced subgraph) a graph  in a hangable supergraph. 

\begin{cor} Every graph  is an induced subgraph of a hangable graph. \end{cor}

\noindent {\bf Proof.} Let $H$ be a graph. We shall prove that $H$ is
an induced subgraph of some hangable graph. We may assume that $H$ is not a hangable graph. Then $H$ is not a complete graph. If no vertex of $H$ is adjacent to every
other vertex of $H$, then
the join $K_1+H$ has exactly one vertex adjacent to every other vertex of $K_1+H$ and it follows from Theorem \ref{hangable-join}\,(2) that  $K_1+H$ is a hangable supergraph
of $H$. Now assume that $H$ has a vertex adjacent to every other vertex of $H$. Let $U$ be the set of all such vertices in $H$. Since $H$ is not a complete graph, the set $V_H-U$ is nonempty and  from the choice of $U$ it follows that $H=H[U]+H[V_H-U]$.
Now no vertex of the graph  $G=(K_1\cup H[U])+H[V_H-U]$ (obtained from $H$ by adding a new vertex and joining it by an edge to every vertex belonging to $V_H-U$) is
adjacent to every other vertex of $G$ and it follows from Theorem \ref{hangable-join}\,(2) that $G$ is a hangable supergraph of $H$.

\section{Open problems}

We conclude this paper with the list of problems.



\begin{Problem} Determine all hangable subgraphs of the cartesian
product $P_m\square P_n$. \end{Problem}

\begin{Problem} Determine all hangable subgraphs of the $n$-cube $Q_n$. \end{Problem}

\begin{Problem}  Determine all graphs $G$ such that $G$ and $\overline{G}$ are hangable graphs.\end{Problem}

\begin{Problem}  Which self-complementary graph $G$ $($such that $G=\overline{G}$$)$ is hangable?\end{Problem}

\begin{Problem} For a connected graph $G$ determine the smallest positive integer $k$ such that the power $G^k$ is a hangable graph.\end{Problem}

\addtolength{\baselineskip}{-1.5mm}


\begin{thebibliography}{9} \bibitem{BuckleyHarary} F. Buckley and F. Harary, Distance in Graphs, Addison-Wesley Publishing Company, Redwood City 1990.
\bibitem{BuckleyLewinter} F. Buckley and M. Lewinter, Graphs with all diametral paths
through distant central nodes, Math. Comput. Modelling 17 (11) (1993), 35-41.
\bibitem{GrossYellen} J. L. Gross and J. Yellen, Graph Theory and Its Applications, Chapman and Hall, Boca Raton 2006.
\bibitem{HandbookProducts} R. Hammack, W. Imrich and S. Klav\v{z}ar, Handbook of Product Graphs, Chapman and Hall, Boca Raton 2011.
\end{thebibliography}
\end{document}